# Ambient-pressure superconductivity onset at 10 K and robust $T_c$ under high pressure in TiNbTaN$_3$ medium-entropy nitride

*Lingyong Zeng,\* Jie Wang, Hongyu Liu, Longfu Li, Jinjun Qin, Yucheng Li, Rui Chen, Jing Song,\* Yusheng Hou,\* Huixia Luo\**

L. Zeng, L. Li, J. Qin, Y. Li, R. Chen, H. Luo
School of Materials Science and Engineering, State Key Laboratory of Optoelectronic Materials and Technologies, Key Lab of Polymer Composite & Functional Materials, Guangdong Provincial Key Laboratory of Magnetoelectric Physics and Devices, Sun Yat-sen University, Guangzhou 510275, China
E-mail: luohx7@mail.sysu.edu.cn (H. Luo)

L. Zeng
Device Physics of Complex Materials, Zernike Institute for Advanced Materials, University of Groningen, 9747 AG Groningen, The Netherlands
E-mail: l.zeng@rug.nl (L. Zeng)

J. Wang, Y. Hou
Guangdong Provincial Key Laboratory of Magnetoelectric Physics and Devices, School of Physics, Sun Yat-sen University, Guangzhou 510275, China
E-mail: houysh@mail.sysu.edu.cn (Y. Hou)

H. Liu, J. Song
*Beijing National Laboratory for Condensed Matter Physics, Institute of Physics,*
Chinese Academy of Sciences, Beijing 100190, China
E-mail: jingsong@iphy.ac.cn (J. Song)

Keywords: superconductivity, medium-entropy nitride, high pressure, new superconductor

**Abstract**

Superconductivity has been one of the focal points in medium and high-entropy alloys (MEAs-HEAs) since the first discovery of the HEA superconductor in 2014. Until now, most HEAs' superconducting transition temperature ($T_c$) has not exceeded 10 K. Here we report the first observation of superconductivity in a bulk medium-entropy nitride (MEN), TiNbTaN$_3$, which shows a $T_c$ of 10 K at ambient pressure. Notably, the electronic specific heat coefficient $\gamma(H)$ exhibits nonlinear H-dependence behavior, which is similar to other well-studied multigap superconductors. Furthermore, TiNbTaN$_3$ exhibits extraordinary pressure resilience, maintaining robust superconductivity under high-pressure conditions. Density functional theory (DFT) calculations indicate that pressure exerts a negligible impact on the electronic structures of TiNbTaN$_3$, thereby corroborating the experimental observations. These findings not only advance our understanding of emergent phenomena in entropy-stabilized nitrides but also establish a new material platform for finding more high-$T_c$ superconductors with combinations of 4$d$/5$d$ transition metal elements and light elements, motivating further investigations into high-entropy functional ceramics for extreme environment applications.

(L. Zeng, J. Wang, and H. Liu contributed equally to this work)

1. Introduction

Superconductivity research holds great significance for societal development, with room-temperature superconductivity being widely regarded as the "holy grail" of condensed matter physics.[1,2] Materials with rich light elements are promising candidates for high critical temperature ($T_c$) superconductivity.[3,4] According to the theoretical perspective based on the Bardeen-Cooper-Schrieffer (BCS) theory, elements with light mass can provide high Debye frequency, thereby contributing to high critical temperatures.[5] Among the most promising material systems, nitride compounds serve as excellent platforms for superconductivity studies. Transition-metal nitrides constitute a large family of materials that hold importance both in fundamental research and technological applications.[6,7] The nitrogen atoms in these superconductors contribute to strong chemical bonding and enhanced electron-phonon coupling, which collectively drive the superconducting behavior.[8] Characterized as hard superconductors, transition-metal nitrides demonstrate particular suitability for extreme condition applications,[9-11] notably since most materials exhibiting superior mechanical hardness tend to be semiconductors or insulators.[12] The fundamental mechanisms underlying superconductivity in transition-metal nitrides have remained unresolved since their discovery.[6] While their superconducting properties, such as energy gaps and upper critical fields, are generally consistent with BCS theory, several early studies have reported anomalous experimental observations that challenge the conventional BCS framework. For instance, measured density of states (DOSs) at the Fermi level combined with estimated electron-phonon coupling strengths appear insufficient to explain the relatively high-$T_c$ observed.[13-15]

Emerging from the concept of medium/high-entropy alloys (MEAs/HEAs), medium/high-entropy nitrides (MENs/HENs) have garnered growing recognition in recent years.[16-18] These multicomponent systems demonstrate superior mechanical properties, oxidation resistance, and thermal stability compared to conventional binary/ternary transition-metal nitrides.[18-21] Significant research momentum has emerged in exploring novel medium/high-entropy superconductors following the groundbreaking discovery of superconductivity in Hf-Nb-Ta-Zr-Ti alloys with a $T_c$ of 7.3 K.[22] Subsequent studies have revealed notable phenomena, including strong electron-phonon coupling, exceptionally high upper critical fields, and topological electronic structures in these complex systems.[23-30] Besides, a record-high $T_c$ of 15.3 K has been observed in body-centered cubic (BCC) MEA superconductor TaNbHfZr at around 70 GPa.[31] The MEA or HEA superconductors that crystallize on the small-cell BCC or CsCl-type lattice have the highest $T_c$. However, all $T_c$s so far are limited to the sub-10 K range under ambient pressure.[31,32] While nitrogen doping strategies are widely implemented to enhance superconducting performance,[33] nitrogen incorporation effects in MEA or HEA thin films have been documented.[34,35] However, no superconducting bulk MEN or HEN has been reported.

Here, we report the discovery of a new bulk MEN superconductor, TiNbTaN$_3$. It shows superconducting behavior around $T_c \approx 10$ K, determined by resistivity, magnetization, and heat capacity measurements on bulk polycrystals. The $T_c$ of TiNbTaN$_3$ is higher than that of other MEA or HEA superconductors (Various MEA and HEA $T_c$ are recorded in Fig. 1(a)[17,22,28-32]). Moreover, by performing high-pressure resistance measurements, we found that the superconductivity in this MEN is robust up to 55 GPa. Our density functional theory (DFT) calculations show that the electronic band structures of TiNbTaN$_3$ near the Fermi level are devoid of topological characteristics. Meanwhile, the band structures and density of states (DOS) of TiNbTaN$_3$ exhibit very weak sensitivity to the pressure, which is consistent with its high-pressure-resistant superconductivity.

2. Results and Discussion

Power x-ray diffraction (PXRD) analysis reveals that TiNbTaN$_3$ adopts a NaCl-type structure with space group $Fm\bar{3}m$ (No. 225), consistent with reported transition-metal nitride systems.[8] As demonstrated in Figure 1b, the Rietveld refinement pattern of TiNbTaN$_3$ exhibits excellent agreement between observed and calculated profiles, confirming phase purity without detectable impurities. All diffraction peaks are indexed with corresponding (hkl) indices, yielding refinement residuals of $R_p$ = 3.00%, $R_{wp}$ = 4.47%, and $\chi^2$ = 5.04 — statistical parameters indicative of high refinement reliability. The cubic lattice parameter determined from refinement is $a = b = c = 4.3266(3)$ Å. Figure 1(c) presents a crystallographic representation of the NaCl-type structure, where Ti, Nb, and Ta are isoelectronically distributed at the cation site while N occupies the anion sublattice. The close-to-atomic energy-dispersive X-ray spectroscopy (EDS) analysis (Figure 1d and Figure S1) confirms microstructural homogeneity and near-equiatomic distribution of metallic constituents (Ti:Nb:Ta ≈ 1:1:1). Moreover, there is no clustering of elements, and each element in the lattice shows distinct and periodic distribution in correspondence with the atomic lattice structure.

Figure 2a shows the temperature-dependent electrical resistivity ($\rho(T)$) of TiNbTaN$_3$ over 1.8 - 12 K, indicating $T_c^{\text{onset}}$ = 10 K and $T_c^{\text{zero}}$ = 9.5 K. External magnetic field effects on the superconducting state are systematically demonstrated in Figure 2b. As expected, the transition becomes broader, and the $T_c$ shifts to the lower temperatures as the applied field is increased. The upper critical field ($\mu_0H_{c2}$) phase diagram was constructed using mid-transition criteria (50 %

ρ$_n$), as plotted in Figure 2c. Experimental μ$_0$H$_{c2}$ data are well-described by the Ginzburg-Landau (GL) formula: $\mu_0 H_{c2}(T) = \mu_0 H_{c2}(0) \times \frac{1-(T/T_c)^2}{1+(T/T_c)^2}$, yielding μ$_0$H$_{c2}$(0) = 8.44 T. Based on BCS theory, the Pauli limiting field for a superconductor can be described as $\mu_0 H_{c2}^P(0) = 1.85 T_c$, that is, $\mu_0 H_{c2}^P(0)$ = 17.58 T for TiNbTaN$_3$, which is larger than the experimental μ$_0$H$_{c2}$(0).

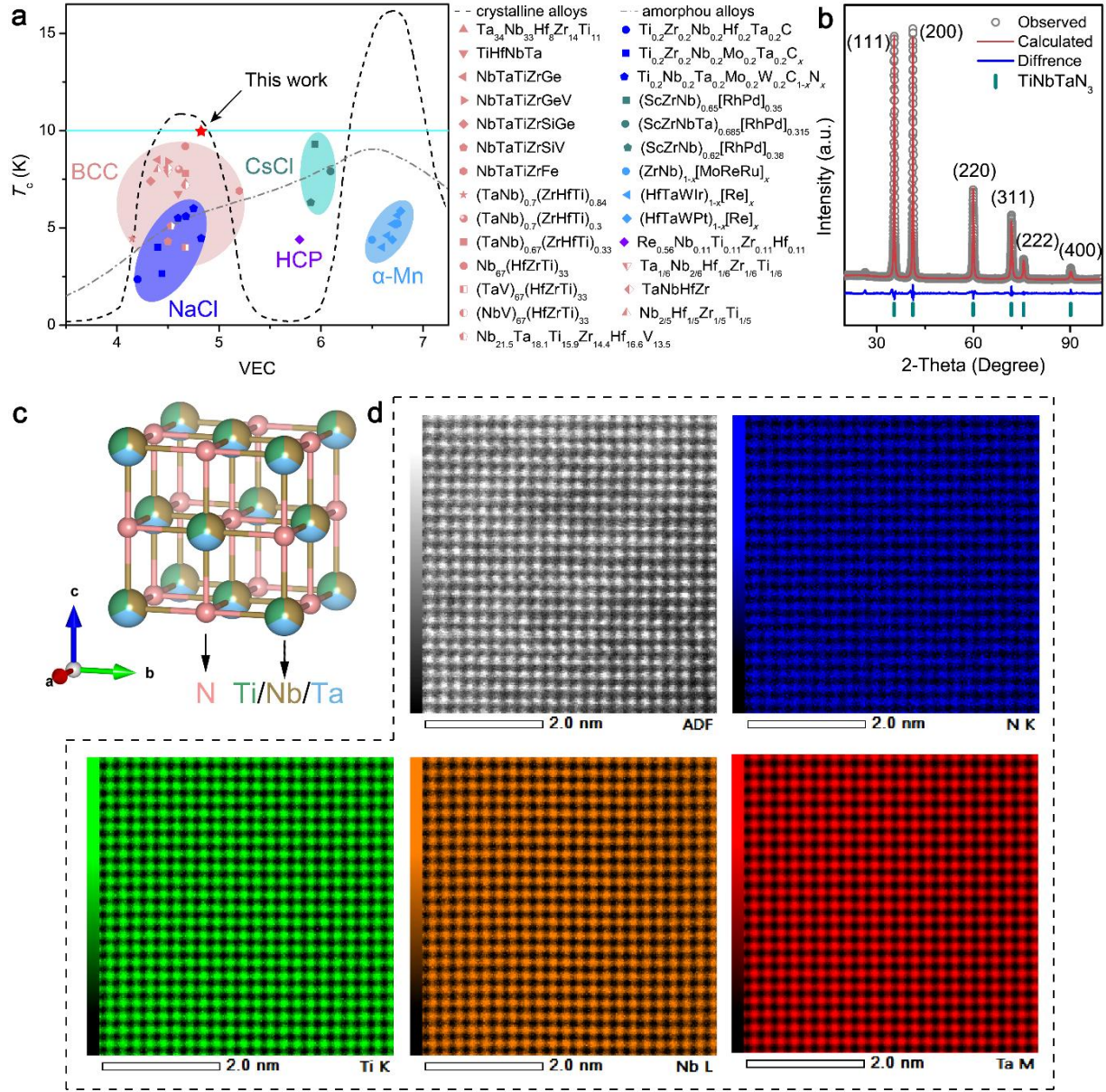

**Figure 1.** Structure characterizations of TiNbTaN$_3$ MEN: a) Valence electron count (VEC) dependence of the $T_c$ for some MEAs-HEAs including TiNbTaN$_3$ MEN under ambient pressure.[17,22,28-32] b) Rietveld refinement and PXRD data. c) The simplified schematic diagram. d) HAADF-STEM images and EDS mappings of TiNbTaN$_3$ MEN.

Bulk superconductivity in MEN TiNbTaN$_3$ is systematically demonstrated through the physical characterization protocol. As shown in Figure 2d, the temperature-dependent zero-field-cooled (ZFC) DC magnetic susceptibility exhibits a sharp superconducting transition at $T_c \approx 9.5$ K under a 2 mT applied magnetic field. To quantify the Meissner response, field-dependent volume magnetization ($M_v$) was measured at 1.8 K within the superconducting state (Fig. S2 [36]). The low-field regime was modeled using the linear relation $M_{fit} = aH + b$, where the slope a determines the demagnetization factor N via the relation: $-a = \frac{1}{4\pi(1-N)}$. This geometry-dependent N value (N = 0.45) was subsequently applied to demagnetization corrections of the susceptibility data in Figure 2d. The corrected diamagnetic susceptibility for TiNbTaN$_3$ achieves $4\pi\chi(1-N) = -1$, thereby confirming bulk superconducting behavior in TiNbTaN$_3$.

Comprehensive vortex dynamics analysis of the MEN TiNbTaN$_3$ was conducted through field-dependent magnetization measurements across the superconducting temperature regime ($1.8 \leq T_c \leq 7.8$ K), as detailed in Figure S2. Following the linear-response methodology discussed previously, the deviation threshold from ideal diamagnetic behavior was quantified via $M_v$-$M_{fit}$ analysis (Figure S3), defining the uncorrected lower critical field $\mu_0 H_{c1}^*$ at each measurement temperature. As demonstrated in Figure 2e, the temperature evolution of $\mu_0 H_{c1}^*$ follows the following formula: $\mu_0 H_{c1}^*(T) = \mu_0 H_{c1}^*(0)(1-(T/T_c)^2)$, where $\mu_0 H_{c1}^*(0)$ represents the extrapolated zero-temperature limit. Experimental data are well-described by this model, yielding $\mu_0 H_{c1}^*(0) = 14.14$ mT. Applying geometry-dependent demagnetization correction (N = 0.45) through the relation: $\mu_0 H_{c1}(0) = \mu_0 H_{c1}^*(0)/(1-N)$, we obtain the intrinsic lower critical field $\mu_0 H_{c1}(0) = 25.71$ mT for TiNbTaN$_3$ MEN.

Based on the results of $\mu_0 H_{c2}(0) = 8.44$ T and $\mu_0 H_{c1}(0) = 25.71$ mT, we can calculate and extract various superconducting parameters for the MEN TiNbTaN$_3$. The GL coherence length, $\xi_{GL}(0) = 62$ Å, can be derived via the formula $\xi_{GL}^2(0) = \frac{\Phi_0}{2\pi\mu_0 H_{c2}(0)}$, where $\Phi_0 = h/2e$ denotes the magnetic flux quantum. Remarkably, $\xi_{GL}(0)$ approaches the range characteristic of heavy-fermion superconductors (e.g., 71 Å in CeIrIn$_5$, 57 Å in CeRhIn$_5$[36]), suggesting a strong pairing potential in this medium-entropy system. Building upon these results, the magnetic penetration depth was derived from: $\mu_0 H_{c1}(0) = \frac{\Phi_0}{4\pi\lambda_{GL}^2(0)} ln \frac{\lambda_{GL}(0)}{\xi_{GL}(0)}$, yielding $\lambda_{GL}(0) = 1414$ Å. Accordingly, from the formula, $K_{GL}(0) = \frac{\lambda_{GL}(0)}{\xi_{GL}(0)}$, we obtain the GL parameter $K_{GL}(0) = 22.8$. This value significantly exceeds the value $1/\sqrt{2}$, unambiguously classifying TiNbTaN$_3$ as a strong type-II superconductor.

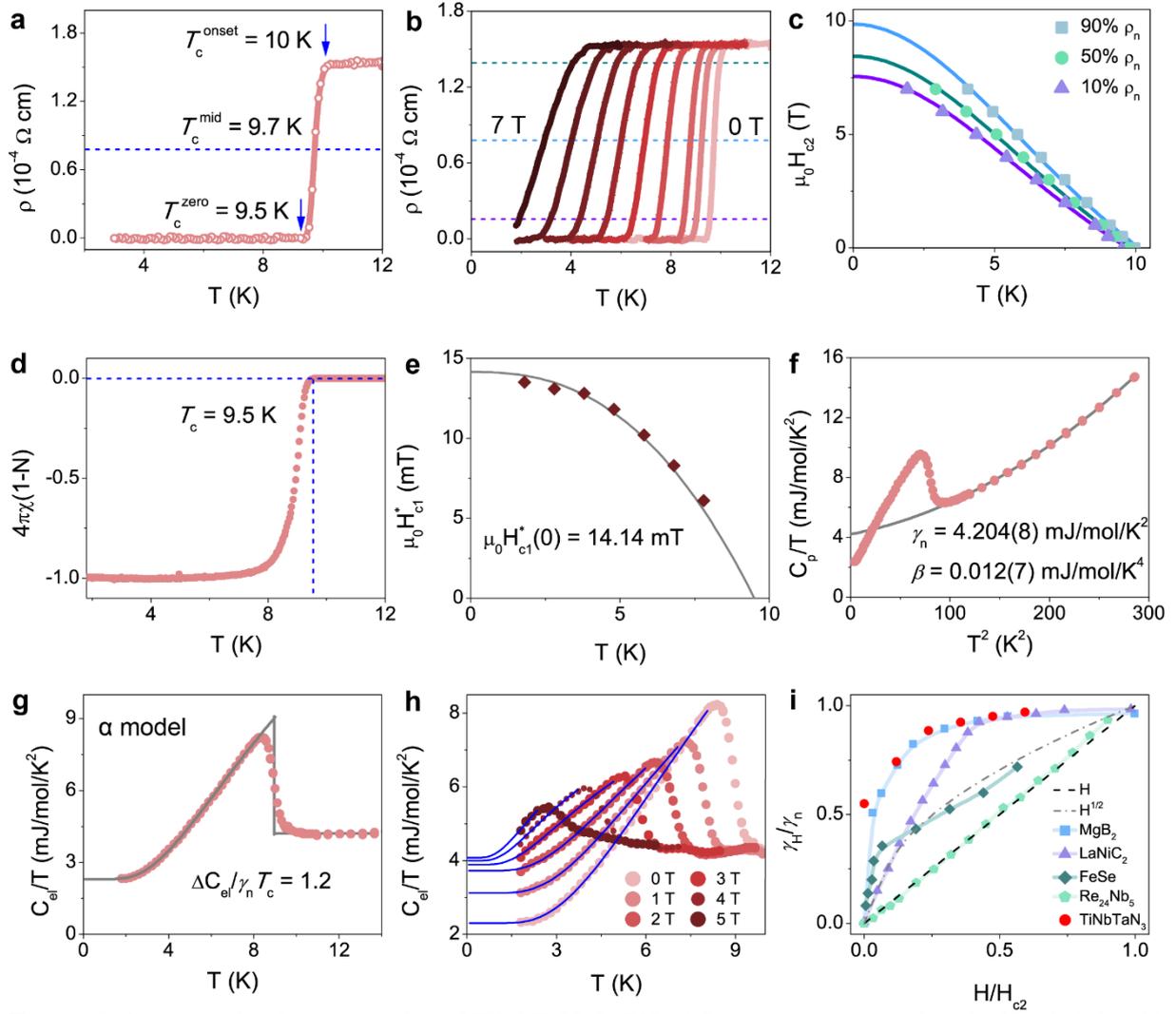

**Figure 2.** Superconducting properties of TiNbTaN$_3$ MEN: a) Low-temperature electrical resistivity. b) Low-temperature electrical resistivity ($\rho$) temperature dependence under varying magnetic fields. c) Temperature-dependent upper critical fields with Ginzburg-Landau (GL) theory fitting. d) Temperature-dependent volume magnetic susceptibility corrected for the demagnetization effect under the ZFC model. e) Lower critical field ($\mu_0 H_{c1}^*$) determination via magnetization hysteresis analysis. f) Temperature-dependent specific heat measured at the static field of 0 T. g) Specific heat jump resulting from the superconducting transition. h) Field-dependent electronic specific heat in fields between 0 and 5 T. i) The normalized specific heat coefficient $\gamma_H/\gamma_n$ vs the reduced magnetic field $H/H_{c2}(0)$. The dash-dotted line represents the dependence expected for an anisotropic gap or a gap with nodes. The dashed line indicates a linear dependence, as predicted for *s*-wave gap structure. The data for the reference are adopted from Ref. [37-40].

To establish bulk superconductivity in MEN TiNbTaN$_3$ beyond transport and magnetic measurements, low-temperature specific heat analysis was conducted under applied magnetic fields (0 - 5 T) as shown in Figure 2f. A sharp anomaly in the 0 applied field heat capacity, corresponding with the emergence of a superconducting state, can be observed beginning at ~

10.0 K, reaching a peak value at ~ 8.4 K, which is consistent with both resistivity and magnetization measurements. In general, the low-temperature specific heat of a solid can be expanded in a power series in temperature: $C_p = \sum a_n T^n$ ($n$ = 1, 3, 5, ⋯), without distinguishing between isobaric and isovolumetric specific heat. However, using only the first two terms cannot describe the data very well, indicating that the simple Debye model cannot describe the material in the low-temperature range. Therefore, we fitted the data with the following formula, $C_p/T = \gamma_n + \beta T^2 + \eta T^4$, where $\beta T^2 + \eta T^4$ is used to express the lattice contribution, and the first term accounts for the contribution from electrons. As displayed in Figure 2f, this formula matches the experimental data well and yields $\gamma_n$ = 4.204(8) mJ/mol/K$^2$, $\beta$ = 0.012(7) mJ/mol/K$^4$, and $\eta$ = 8.46 × 10$^{-5}$ mJ/mol/K$^6$. To extract the value of the specific heat jump, an equal-area entropy construction was employed in Figure 2g. The normalized jump ($\Delta C_{el}/\gamma T_c$) is determined as 1.2, close to the expected value of 1.43 for BCS superconductivity in the weak coupling limit, suggesting a bulk transition to the superconducting state. Moreover, the Debye temperature $\Theta_D$ can be calculated according to $\Theta_D = (12\pi^4 nR/5\beta)^{1/3}$, where $n$ is the number of atoms per formula unit and R represents the gas constant. The $\Theta_D$ of TiNbTaN$_3$ is found to be 673 K. Based on the McMillan formula, we estimated the electron-phonon coupling to be $\lambda_{ep}$ = 0.61, using $\lambda_{ep} = \frac{1.04 + \mu^* \ln\left(\frac{\Theta_D}{1.45 T_c}\right)}{(1 - 0.62\mu^*) \ln\left(\frac{\Theta_D}{1.45 T_c}\right) - 1.04}$, where $\mu^*$ is a Coulomb pseudo-potential and has a typical value of 0.13 for intermetallic superconductors.[29,41,42]

In light of the observed flattening behavior at low temperatures, we have conducted an analysis of the $C_{el}/T$ data utilizing the $\alpha$ model (see Figure 2g), which has been derived from BCS theory but modified to consider the multiple bands, gap anisotropy, strong coupling, and other aspects.[43] In zero fields, the extrapolated value of $C_{el}/T$ is finite and nonzero, which may be designated as the residual electronic specific heat coefficient ($\gamma_r$). From the analysis of magnetic susceptibility and specific heat jump data, we know that the superconducting phase of this sample is close to 100 %. Therefore, such a fraction of $\gamma_r$ is not expected, and the $\gamma_r$ cannot be due to the non-superconducting phase in the sample. In real materials, the existence of defects, impurities, and disorders is unavoidable. These scattering effects can substantially impact quasiparticle excitations in superconductors. Medium- and high-entropy compounds are a class of highly disordered materials. Consequently, this finite $\gamma_r$ in zero fields in TiNbTaN$_3$ MEN may be due to non-superconducting quasiparticles caused by disorder. The residual contributions to the superconductivity at low temperatures arose from a strong disorder-induced localization of the uncondensed quasiparticles, which do not participate in the superconducting transition.

To get more information about the superconducting gap, we also measured the low-temperature specific heat at various magnetic fields, as shown in Figure 2h. By increasing the magnetic field, the magnitude of the specific heat jumps at $T_c$ decreases, and the specific heat coefficient $\gamma(H)$, which is obtained by linear extrapolations of the low-temperature specific heat curves, increases. And we examined the field dependence of $\Delta\gamma = \gamma(H) - \gamma(0)$; the obtained results are shown in Figure S4. Considering the apparent deviation of $\Delta\gamma$ from linear field dependence, we fitted the data with a power law behavior, $\Delta\gamma = R(\mu_0H)^n$, where R and n are the fitting parameters. In a conventional nodeless $s$-wave superconductor, the $\Delta\gamma$ is expected to vary linearly with the applied field, while $d$-wave superconductors display a square root dependence on the field.[44-46] An intermediate behavior is generally interpreted in iron-based superconductors as due to the presence of gaps with different amplitudes.[47,48] The obtained exponents n of TiNbTaN$_3$ at 0 K and 1.8 K are 0.41 and 0.68, respectively. Additionally, the nonlinear contribution of $\Delta\gamma$ has also been observed in many other superconductors exhibiting multigap features. Figure 2i shows the normalized values $\gamma_H/\gamma_n$ vs the reduced magnetic field $H/H_{c2}(0)$ of the typical multiband superconductors. TiNbTaN$_3$ instead exhibits features similar to other well-studied multigap superconductors,[37-40] e.g., MgB$_2$, LaNiC$_2$, and FeSe, although the slopes of $\gamma_H(H)$ close to zero field are different, reflecting the different magnitudes and weights of the smaller gap. Therefore, the nonlinear contribution of $\Delta\gamma$ in TiNbTaN$_3$ may be ascribed to multiband effects. Of course, it is necessary to confirm the unconventional superconductivity in TiNbTaN$_3$ employing other experiments, such as the thermal conductivity, angle-resolved photoemission spectroscopy (ARPES), muon spin rotation (µSR), and nuclear magnetic resonance (NMR) measurements at lower temperatures in future studies.

Figure 3a shows the temperature dependence of electrical resistance measured in the pressure range of 2.8 - 54.5 GPa for TiNbTaN$_3$ MEN. The superconducting transition of the sample under various pressures is observed to be sharp, and the zero-resistance state remains robust across all applied pressures. With these results, the electronic P-T phase diagram for TiNbTaN$_3$ MEN is established, as shown in Figure 3b. It is found that the $T_c$ is rather robust against pressure, with a slight $T_c$ variation of 1 K within 50 GPa, similar to high-entropy carbides and some HEAs,[29,49] unlike typical superconductor (Pb) ($\approx$ 5 K)[50] or unconventional superconductors (cuprate, iron-based superconductors, etc.) ($\approx$ 20 K).[51,52] The robustness of superconductivity in this MEN superconductor could be related to its outstanding structural stability under pressure. This makes TiNbTaN$_3$ a superconductor, also promising candidates for new applications under extreme conditions. To further probe the evolution of superconductivity in TiNbTaN$_3$ under pressure, we examined the effect of applied magnetic fields. Figure 3c shows the superconducting transition under different magnetic fields at 54.5 GPa. Here we also used the criteria of 50 % of normal-state resistivity for determining $T_c$(H), from which the $\mu_0H_{c2}(0)$ is calculated to be 6.71 T (seen in Figure 3d). Compared to the $\mu_0H_{c2}(0)$ of normal ambient, the upper critical field is reduced a little after pressurization.

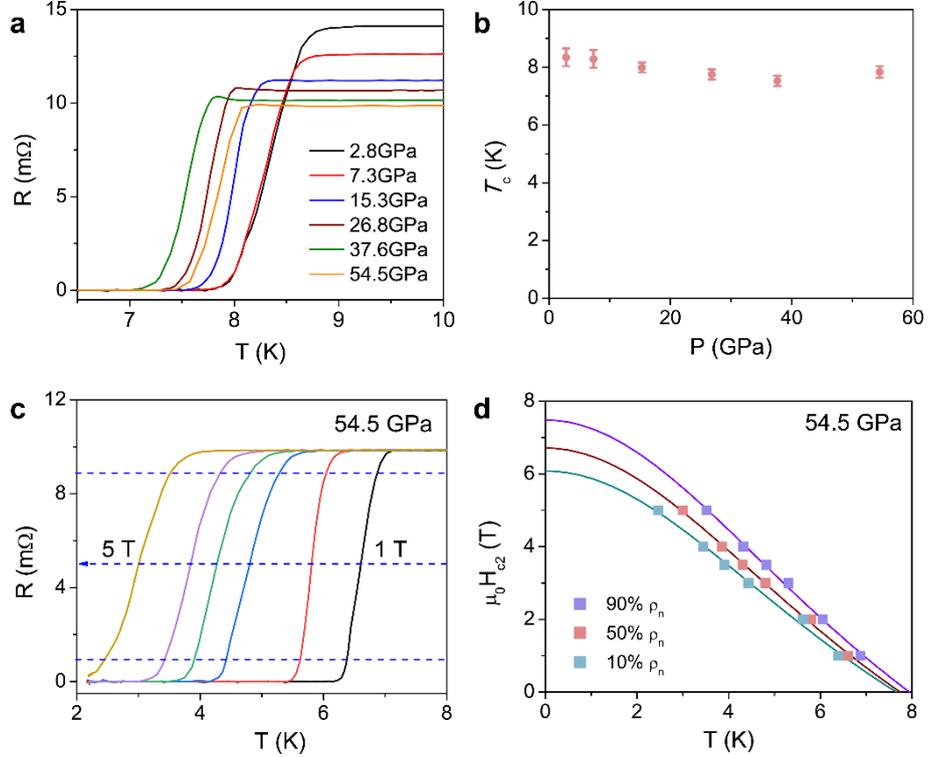

**Figure 3.** Transport properties of the TiNbTaN$_3$ under high pressure: a) Superconducting transition at pressure up to 54.5 GPa. b) The electronic P-T phase diagram. c) Superconducting transition under various magnetic fields at 54.5 GPa. d) Upper critical field as a function of temperature at 54.5 GPa.

To gain more insight into TiNbTaN$_3$, here we perform DFT calculations to study its electronic properties. Considering that TiNbTaN$_3$ has a highly symmetric NaCl-type structure, which may exhibit topological features, such as topological nodal points and lines predicted in CuN [53,54], we first examine its electronic band structures as shown in Figure 4a-b. In CuN, a Weyl point was identified along the high-symmetry path X − U. Due to symmetry constraints, there were a total of four Weyl points within the $k_y = -2\pi/a$ plane. These Weyl points were safeguarded by the time-reversal symmetry and C$_2$ rotational symmetry, rendering them robust in the presence of spin-orbit coupling (SOC).[54] However, as shown in Figure 4a, no similar situation is found along the X-U path in the TiNbTaN$_3$ band structure when SOC is not considered. It is worth noting that several crossing points are observed along the Γ − X and K − Γ paths. This motivates us to delve deeper into band structure with SOC considered, as depicted in Figure 4b. It can be seen that the original crossing points open into noticeable gaps. On the Γ − X path, a crossing point is detected, which is located 0.84 eV away from the Fermi level. Additionally, in CuN, the Γ point was a quadratic contact point, whereas in TiNbTaN$_3$, the Γ point lacks such a characteristic. Therefore, our calculations indicate that TiNbTaN$_3$ does not exhibit apparent topological properties in its band structure around the Fermi level.

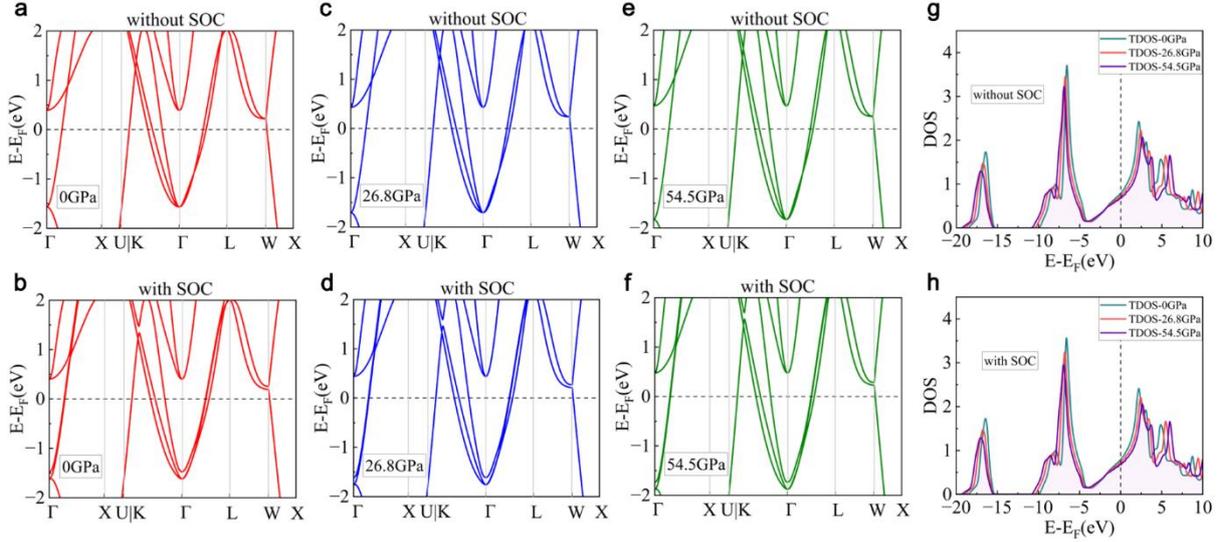

**Figure 4.** DFT calculations of the TiNbTaN$_3$ under typical pressures: The electronic band structures without SOC at a) 0 GPa, c) 26.8 GPa, e) 54.5 GPa, and with SOC at b) 0 GPa, d) 26.8 GPa, f) 54.5 GPa. The DOS g) without SOC and h) with SOC under 0, 26.8, 54.5 GPa.

Given that TiNbTaN$_3$ possesses superconductivity that remains largely resilient to high pressure, we explore the effect of pressure on its band structures and DOS by performing calculations at the experimentally relevant pressures of 26.8 and 54.5 GPa. The optimized lattice constants under different pressures are presented in Table S1. As shown in Figure 4c-f, with the applied pressure increasing, the band structures remain chiefly unaltered, manifesting only minor energy shifts near the Γ point, in general. Similar to the case at 0 GPa, when SOC is not considered, the band structures under pressures of 26.8 and 54.5 GPa reveal a couple of band crossings along the Γ − X and K − Γ paths. Nevertheless, upon incorporating SOC, these points are transformed into distinct gaps. This illustrates that pressure almost does not affect the band structure of TiNbTaN$_3$. Then, we calculate the DOS of TiNbTaN$_3$ both with and without SOC under 0, 26.8, and 54.5 GPa in Figure 4g-h. We see that the SOC has a minimal impact on the electronic DOS of TiNbTaN$_3$. The Fermi level is situated below the flank of the DOS peak, while a more pronounced peak exists beneath the Fermi level, but the energy gap between them is excessively large. By comparing the DOS under various pressures, we observe that pressure exerts a slight influence on the overall features of the DOS. It predominantly results in a diminution of the peak intensities, whereas the DOS near the Fermi level displays negligible variation in response to external pressure. The calculated band structures and DOS under 26.8 and 54.5 GPa suggest that the pressure has a very weak effect on the electronic properties of TiNbTaN$_3$. This invariance likely underpins the robust superconductivity of TiNbTaN$_3$ at high pressure, aligning well with the experimental results.

3. Conclusion

We present the successful synthesis and comprehensive characterization of a novel MEN superconductor, TiNbTaN$_3$. Our studies show that TiNbTaN$_3$ is a type-II superconductor with $T_c^{\text{onset}}$ = 10 K, $T_c^{\text{zero}}$ = 9.5 K, $\mu_0 H_{c2}(0)$ = 8.44 T, and $\mu_0 H_{c1}(0)$ = 25.71 mT. Notably, the electronic specific heat coefficient $\gamma(H)$ exhibits nonlinear H-dependence behavior, which is similar to other well-studied multigap superconductors, though definitive confirmation requires phase-sensitive experimental verification (e.g., Josephson junction spectroscopy or spin-polarized STM). This anomalous behavior suggests potential unconventional though definitive confirmation requires phase-sensitive experimental verification in the future. Furthermore, the observed high-pressure resilience of superconductivity ($\Delta T_c$ < 1 K up to 54.5 GPa) combined with excellent mechanical properties positions MENs/HENs as promising candidate materials for extreme-condition quantum devices. Beyond technological implications, this emergent material platform opens new avenues for exploring entropy-stabilized superconducting states through strategic compositional engineering of 4$d$/5$d$ transition-metal, including light metals (e.g., N/C/B) systems. This is certainly an interesting idea, which will open new avenues for exploring new high-$T_c$ superconductors with combinations of 4$d$/5$d$ transition metal elements and light elements.

4. Experimental Section

*Sample Characterization*: The TiNbTaN$_3$ compound was synthesized via spark plasma sintering (SPS). High-purity commercial TiN (99.9 %, Macklin), NbN (99.9 %, Macklin), and TaN (99.9 %, Macklin) powders served as precursor materials. Following stoichiometric proportioning, the raw powders were homogenized through ball milling in anhydrous ethanol using ZrO$_2$ grinding media. The blended powder was subsequently loaded into a graphite die assembly within the SPS chamber for consolidation. Sintering was performed at 2000 ℃ for 15 minutes with a heating rate of 100 ℃/min under a 1 atm nitrogen atmosphere. Structural verification was conducted through room-temperature powder X-ray diffraction (PXRD) using a Rigaku MiniFlex diffractometer operating at 1 ℃/min scanning rate. Atomic-resolution HAADF-STEM images and EDS mappings were collected on a JEM-ARM200F NEOARM microscope equipped with a JEOL's EDS detector at an acceleration voltage of 200 kV. Resistivity, magnetization, and heat capacity measurements were carried out in a Quantum Design PPMS-14T.

*High-pressure measurements*: High pressures were generated using a diamond anvil cell (DAC) made of CuBe alloy, equipped with two opposed diamond anvils featuring 0.3 mm culets. A thin square-shaped high-entropy ceramic (HEC) sample was placed atop four thin gold leads for four-point dc electrical resistivity measurements. A cubic boron nitride (cBN)-epoxy mixture was used to insulate the T301 stainless steel gasket, and NaCl served as the pressure-transmitting medium. Pressure calibration was performed using a ~10 μm ruby sphere as a pressure manometer. The high-pressure resistivity techniques were reported in detail elsewhere.[55,56]

*DFT details*: The first-principles calculations based on the density functional theory (DFT) were implemented with the Vienna *Ab initio* Simulation Package (VASP) at the level of the generalized gradient approximation.[57-60] The crystal structure of TiNbTaN$_3$ was modeled by the virtual crystal approximation (VCA) method.[61-63] Given that Ta, Ti, and Nb resided in the same or adjacent groups of the periodic table and exhibited similar elemental properties, the VCA was well-suited for this scenario.[64,65] To further minimize potential inaccuracies, we used projector-augmented-wave pseudopotentials and ensured that the atoms located at the same positions had similar valence electrons, namely Ti$3s^23p^63d^4$, Nb$4s^24p^64d^5$, Ta$5s^25p^65d^5$.[66] The energy cutoff of the plane-wave expansion was set to be 450 eV. To ensure the accuracy of our calculations, we also tested the cut-off energies of 500, 550, and 600 eV. As shown in Fig. S5, the calculated band structures and DOSs indicate that the effect of cut-off energy on our results is negligible, thereby validating the choice of the cut-off energy of 450 eV in our study. For the non-pressurized TiNbTaN$_3$, the experimental structure was adopted to calculate its electronic properties. For TiNbTaN$_3$ under pressure, we fully optimized its lattice constant and atomic positions with a k-mesh of 25×25×25 until the force on each atom was less than 0.01eV/Å. In the self-consistent calculations, a denser k-point grid with 39×39×39 was employed.

Acknowledgments

L. Zeng, J. Wang, and H. Liu contributed equally to this work. This work is supported by the National Natural Science Foundation of China (12404165, 12274471, 11922415, 12474247, and 92165204), Guangdong Basic and Applied Basic Research Foundation (Grant No.

2025A1515010311), Guangzhou Science and Technology Programme (No. 2024A04J6415), the State Key Laboratory of Optoelectronic Materials and Technologies (Sun Yat-Sen University, No. OEMT-2024-ZRC-02), Key Laboratory of Magnetoelectric Physics and Devices of Guangdong Province (Grant No. 2022B1212010008), and Research Center for Magnetoelectric Physics of Guangdong Province (2024B0303390001). Lingyong Zeng acknowledges the Postdoctoral Fellowship Program of CPSF (GZC20233299) and the Fundamental Research Funds for the Central Universities, Sun Yat-sen University (24qupy092). J. S. is thankful to the National Natural Science Foundation of China (12204514) and the National Key Research and Development Program of China (2023YFA1406000).

# Supporting Information

**Ambient-pressure superconductivity onset at 10 K and robust $T_c$ under high pressure in TiNbTaN$_3$ medium-entropy nitride**

*Lingyong Zeng,\* Jie Wang, Hongyu Liu, Longfu Li, Jinjun Qin, Yucheng Li, Rui Chen, Jing Song,\* Yusheng Hou,\* Huixia Luo\**

L. Zeng, L. Li, J. Qin, Y. Li, R. Chen, H. Luo
School of Materials Science and Engineering, State Key Laboratory of Optoelectronic Materials and Technologies, Key Lab of Polymer Composite & Functional Materials, Guangdong Provincial Key Laboratory of Magnetoelectric Physics and Devices, Sun Yat-sen University, Guangzhou 510275, China
E-mail: luohx7@mail.sysu.edu.cn (H. Luo)

L. Zeng
Device Physics of Complex Materials, Zernike Institute for Advanced Materials, University of Groningen, 9747 AG Groningen, The Netherlands
E-mail: l.zeng@rug.nl (L. Zeng)

J. Wang, Y. Hou
Guangdong Provincial Key Laboratory of Magnetoelectric Physics and Devices, School of Physics, Sun Yat-sen University, Guangzhou 510275, China
E-mail: houysh@mail.sysu.edu.cn (Y. Hou)

H. Liu, J. Song
*Beijing National Laboratory for Condensed Matter Physics, Institute of Physics,*
Chinese Academy of Sciences, Beijing 100190, China
E-mail: jingsong@iphy.ac.cn (J. Song)

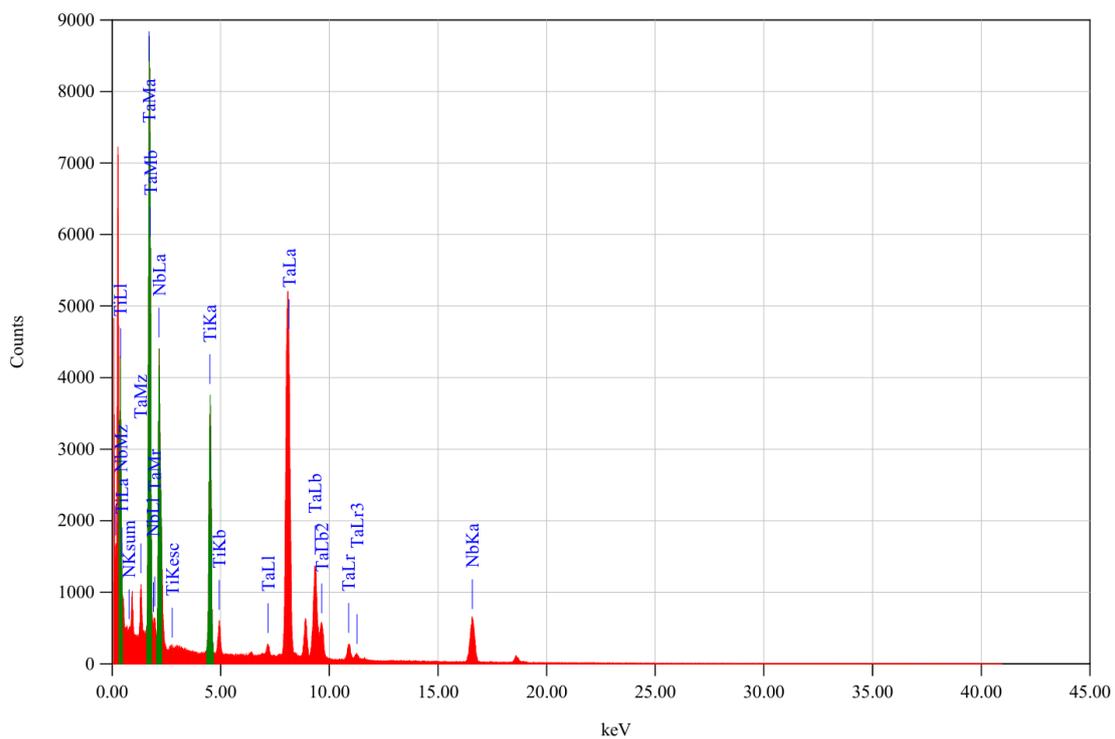

**Figure S1.** EDS spectrum of TiNbTaN$_3$ MEN.

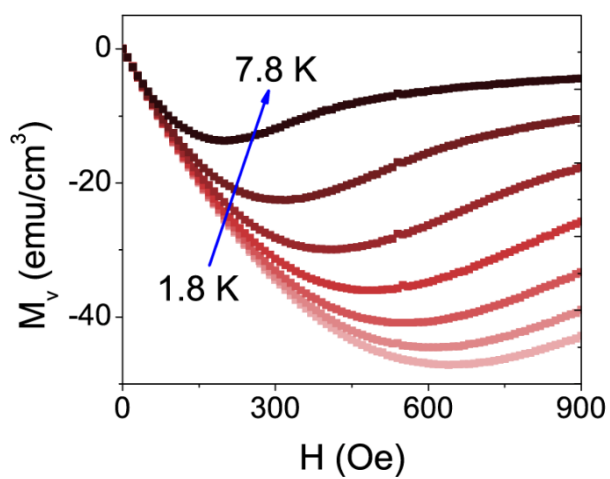

**Figure S2.** Isothermal magnetization curves (M vs. H) in the regime (0 - 900 Oe; T = 1.8 - 7.8 K).

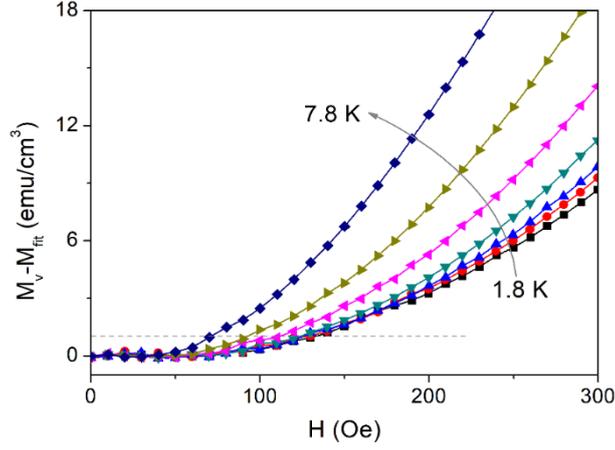

**Figure S3.** The difference between $M_v$ and $M_{fit}$ at 0 - 300 Oe under several temperatures for TiNbTaN$_3$ MEN.

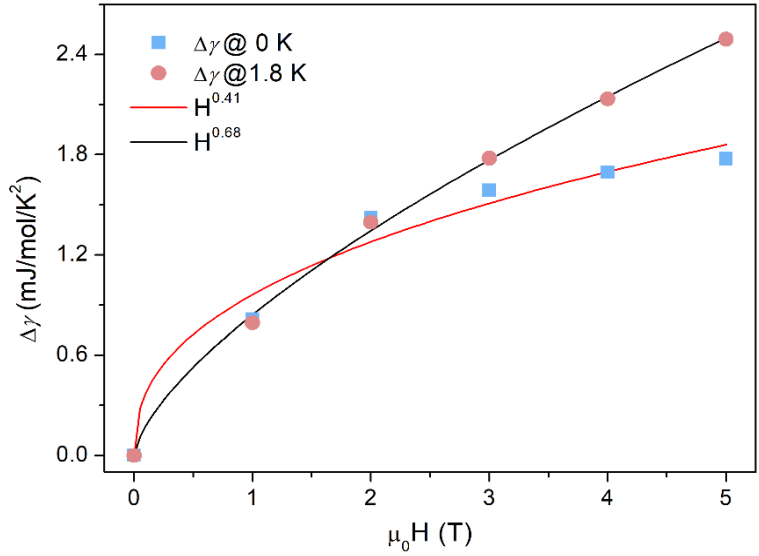

**Figure S4.** The field dependence of reduced electronic specific heat coefficient $\Delta\gamma$ at 0 K and 1.8 K.

**Table S1.** The optimized lattice constants and internal coordinates of TiNbTaN$_3$ under different pressures.

| Pressure (GPa) | 0 | 26.8 | 54.5 |
| --- | --- | --- | --- |
| Lattice constants (Å) | 4.3266(3) (*exp.*) | 4.2341 | 4.1596 |
| Internal coordinates | Ti/Nb/Ta (0, 0, 0)<br>N (0.5, 0.5 ,0.5) | Ti/Nb/Ta (0, 0, 0)<br>N (0.5, 0.5 ,0.5) | Ti/Nb/Ta (0, 0, 0)<br>N (0.5, 0.5 ,0.5) |

The lattice constant used at 0GPa is based on the experimental value. The internal coordinates remain unchanged after structrual relaxtions under different pressures due to the high symmetry of the crystal structure of TiNbTaN$_3$.

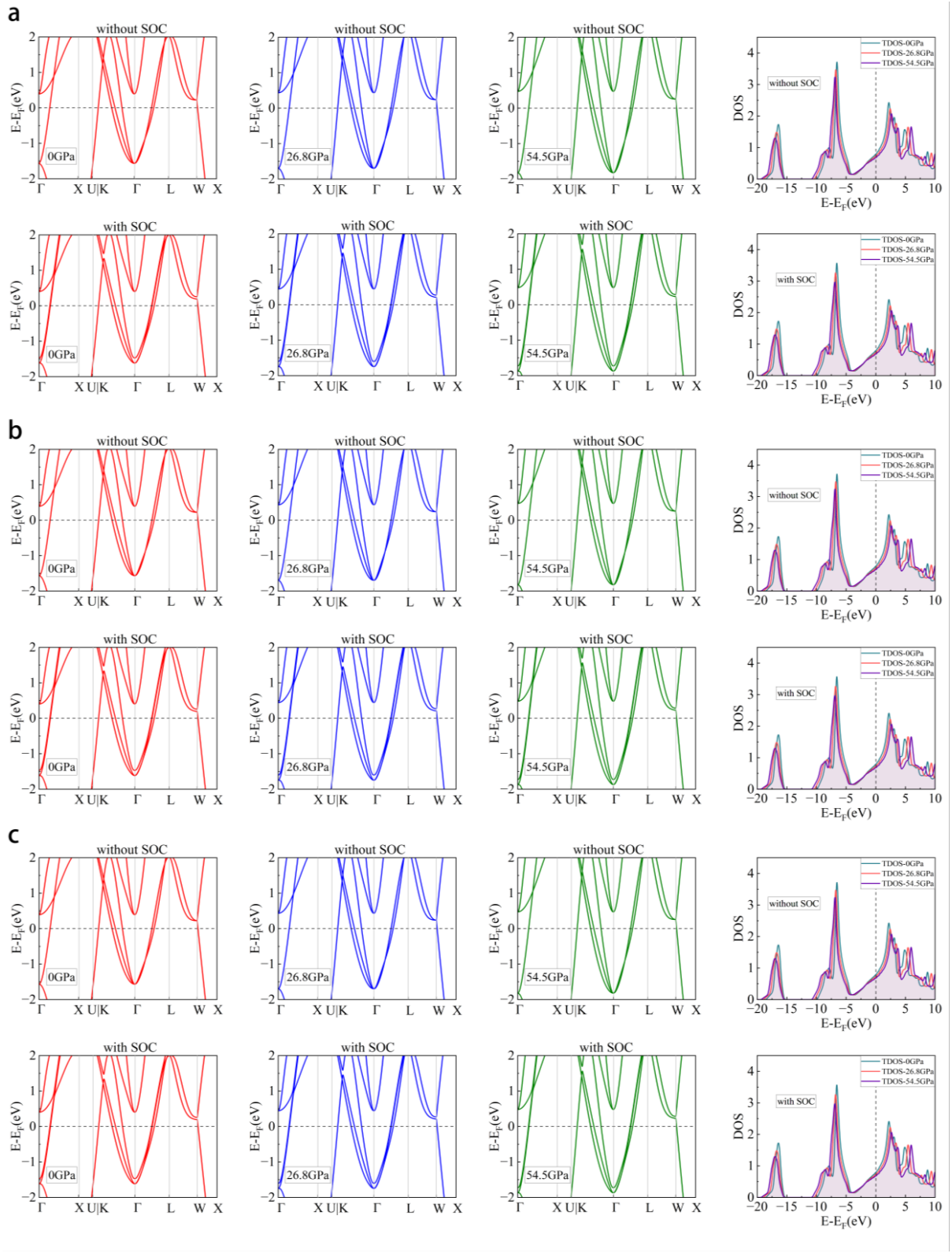

**Figure S5.** DFT calculated band structures and DOS (with and without SOC) of TiNbTaN$_3$ under different pressures using the cut-off energies of a) 500 eV, b) 550 eV, c) 600 eV.